\documentclass[12pt,aps,prb,floats,eqsecnum,graphicx,epsfig,euscript,amsmath,amssymb,onecolumn]{revtex4-1}
\usepackage[dvips]{graphicx}
\usepackage{epsfig}
\DeclareMathAlphabet\EuScript{U}{eus}{m}{n} \SetMathAlphabet\EuScript{bold}{U}{eus}{b}{n}

\renewcommand{\min}{\mathop{\rm min}\nolimits}

\def\lapprox{\,\raise0.4ex\hbox{$<$}\kern-0.8em\lower0.7ex\hbox{$\sim$}\,}
\def\gapprox{\,\raise0.4ex\hbox{$>$}\kern-0.8em\lower0.7ex\hbox{$\sim$}\,}
\begin{document}
\bibliographystyle{prsty}
\title{Competing Hyperfine and Spin-Orbit Couplings: Spin Relaxation in a Quantum Hall Ferromagnet}

\author{S. Dickmann$^{1,2}$ and T. Ziman$^{2}$}
\affiliation{$^{1}$Institute for Solid State Physics of RAS, Chernogolovka 142432, Moscow
District, Russia.\\
$^{2}$CNRS and Institut Laue Langevin, 6 rue Jules Horowitz, BP 156 - 38042, Grenoble, France}

\date{\today}

\begin{abstract}
\vspace{0.mm}
Spin relaxation in a quantum Hall  ferromagnet, where filling is $\nu=1, 1/3, 1/5,...$,
can be  considered in terms of spin wave
annihilation/creation processes. Hyperfine coupling with the nuclei of the GaAs matrix provides  spin non-conservation in the two-dimensional electron gas and determines
spin relaxation in the quantum Hall system. This mechanism competes with  spin-orbit coupling
channels of  spin-wave decay and can even  dominate in
a  low-temperature regime where $T$ is much smaller than the Zeeman gap.
In this case the spin-wave relaxation  process occurs non-exponentially with
time and does not depend on the temperature. The competition  of  different relaxation channels results in crossovers in the dominant mechanism, leading to   non-monotonic behavior of the characteristic relaxation time with the magnetic field.
We predict that the relaxation times should reach maxima at $B\simeq \!18\,$T in the $\nu\!=\!1$ Quantum Hall system and at $B\simeq \!12\,$T for that of  $\nu\!=\!1/3\,$. We estimate these times as $\sim\!10\,-\,30\,\mu$s and $\sim\!2\,-\,5\,\mu$s, respectively.

\noindent PACS numbers 73.43.Lp, 78.67.De, 73.21.Fg
\end{abstract}
\maketitle

\bibliographystyle{prsty}

\section{Introduction}

The two-dimensional (2D) electron gas has been
intensively studied for several decades. The interest is stimulated    by the  clear  manifestations of strong electron
correlations including quantum phase
transitions (like Wigner crystallization${\,}$\cite{Wigner}) and,
in the presence of a strong perpendicular magnetic field, to various
features in electron transport gathered under the name
``fractional quantum Hall effect''.\cite{FQHE} Transport
phenomena, although of  paramount significance for applications,
provide  only
indirect information on such fundamental characteristics as
quantum states and the energy spectrum, where optical techniques
give much more immediate information. In particular, Raman scattering,
starting from the pioneering works of A. Pinczuk {\em et al},\cite{Pi}
has been  successfully used to study collective
excitations in two-dimensional electron gas created in
semiconductor heterostructures and quantum wells (see also Ref.
\onlinecite{ku06} and references therein). The position and
intensities of corresponding Raman or luminescence lines
can yield information on the energy and oscillator strengths of the
excited states. Meanwhile an important characteristic of
such excitations is also the life-time. This may be estimated by observation of the resonance line widths: for example the spin wave life-time was deduced from the  observed width of the ESR resonance lines.\cite{ne10}. Microwave and optical
linewidths are not, however,  directly related to the life-time, and  usually provide only a very rough lower bound for this quantity. In consequence one is forced to use combined experimental methods including a time-resolved technique (see, e.g., Ref. \onlinecite{zh93}).
Despite these experimental difficulties, growing interest in the problem
of excitation life-times in a two-dimensional electron gas has been observed in recent years.
One should mention, for example, recent experimental works on the observation of the spin
relaxation in a polarized two-dimensional electron gas based on the Kerr rotation effect.\cite{fu08}

We study in this
work the so-called quantum Hall ferromagnet where all two-dimensional electron gas
electrons of the upper,  partially  filled Landau level,
are in the ground state, with spins aligned along the magnetic field. This
state obviously arises at odd integer fillings:
$\nu=1,3,...$.\cite{by81} In addition, experiments and
semi-phenomenological theories show that at some fractional
fillings, namely at $\nu=1/3,1/5,...$, electrons in the ground
state occupy only one spin sub-level, and thereby the fractional
quantum Hall ferromagnet state is also realized.\cite{gi85,lo93,kukush06,va09,ku09}
The quantum Hall ferromagnet possesses a macroscopically large spin ${\vec S}$
oriented in the direction of the field ${\vec B}$ due to negative
$g$-factor in GaAs structures.
The spin wave in the quantum Hall ferromagnet may be defined as a purely electronic collective
excitation within the Landau level which corresponds to a change of the spin
numbers by one:
\begin{equation}\label{deltaS}
   \delta S=\delta S_z=-1\,,
\end{equation}
and does not alter the spin {\it orientation} of the system. (Another
possible excitation in the quantum Hall ferromagnet is a Goldstone mode representing a
deviation of ${\vec S}$ from the ${\vec B}$ direction which does not change
the $S$ number;\cite{di04} then the microscopic excitation would be
a ``zero spin exciton'' corresponding to the spin change $\delta S_z=-1$, but
$\delta S=0$.) This spin wave is also called the spin exciton, because this excitation promotes an electron to another spin
sub-level of the same Landau level and thus an effective hole appears in
the initial sub-level. Every spin exciton possesses
energy$\,$\cite{by81,lo93}
\begin{equation}\label{ex_energy}
  E_{\rm x}=\epsilon_{\rm Z}+{\cal E}_q,
\end{equation}
where $\epsilon_{\rm Z}=|g|\mu_BB$ is the Zeeman gap ($g\approx
-0.44$ in a GaAs structure); ${\cal E}_q$ is the spin exciton Coulomb
correlation energy depending on the 2D wave vector modulus $q$. For the rest of the paper  it will be
sufficient to consider only long wave excitations,
$q\ll 1/l_B$ ($l_B$ is the magnetic length), for which the
spectrum is quadratic:
\begin{equation}\label{spectrum}
{\cal E}_q\approx q^2l_B^2/2M_{\rm x}.
\end{equation}
Here the spin-exciton mass $M_{\rm x}$ has the dimensionality of
inverse energy.\cite{by81,lo93} This quantity has recently
been measured experimentally, for  $\nu=1$ \cite{kukush09,ga08} and
$\nu=1/3$ fillings.\cite{kukush06}

If there are an excessive number of spin excitons compared with
equilibrium, then the spin relaxation reduces to an elementary
process of spin exciton annihilation. The spin numbers are changed
in accordance with Eq. (\ref{deltaS}), where the energy of the
annihilated excitation can be transferred to the emitted acoustic
phonon or to another exciton due to the spin-exciton - spin-exciton scattering. Any spin exciton
relaxation channel is thus determined by two necessary conditions:
by the availability of an interaction that does not conserve the   spin of the electron gas, and
by a mechanism of energy dissipation making the relaxation
process irreversible. Until now, spin-orbit coupling has been
assumed to be the cause of the spin non-conservation (see, e.g.,
Refs. \onlinecite{di04,di09} and the works cited therein). Indeed these
spin-orbit relaxation channels are certainly dominant under the usual experimental
conditions, where $T\sim 1\,$K and $1<B<10\,$T. The corresponding
calculations are in satisfactory
agreement with the available experimental data. Here
 we shall extend the study of spin
relaxation channels to include spin non-conservation
by  the hyperfine coupling to nuclei of the GaAs matrix. This has been considered previously$\,$\cite{di11} only for the case of the Goldstone mode $q\equiv 0$; here we consider non-zero,  but small $q$.
Our analysis shows that one mechanism in particular, relating to the spin-exciton - spin-exciton scattering process, should be
taken into account, if $T\lapprox 0.1\,$K and magnetic fields $B \gapprox 10\,$T.\cite{di11} (Specifically, the necessary condition is $T\ll \epsilon_{\rm Z}$.)
To see this clearly  we will analyse  the spin-orbit relaxation channels -- two of them can compete with the hyperfine coupling relaxation in the same  region of temperature and magnetic fields.

It should  also be noted that the spin relaxation processes proceed
much more slowly than other two-dimensional electron gas plasma relaxations unrelated to
a spin change. This means that in any case  an elementary spin exciton annihilation/creation process may be studied as a transition (induced by a perturbation) from an initial { eigen} state $|{i}\rangle$ to a final eigenstate $|{f}\rangle$; i.e. the hyperfine coupling relaxation mechanisms are governed, like the spin-orbit coupling relaxation,\cite{di04,di09,di96,2di96,di99} by the Fermi golden rule probability
\begin{equation}\label{FGR}
w_{fi}=(2\pi/\hbar)|{\cal M}_{fi}|^2\delta(E_f-E_i)\,,
\end{equation}
where ${\cal M}_{fi}$ is a relevant matrix element.

In principle, the hyperfine coupling effects are weak. The spin-orbit coupling and the hyperfine coupling
both have relativistic origins: the former is of the first order, but the
latter represents  the second order relativistic
correction to the Hamiltonian. However, the hyperfine interaction has
some essential properties different from those of the spin-orbit coupling. These
substantially change kinematic conditions of the spin exciton scattering and the dissipation mechanisms where one of spin excitons annihilates. We shall see
that (i) first, the hyperfine coupling does not
conserve total momentum of the electron system, and this feature leads
to extension of the phase volume for the spin-exciton - spin-exciton and spin-exciton - phonon scatterings; (ii) second, the spin-flip
process governed be the hyperfine interaction does not require a
virtual promotion of an electron to another Landau level (this promotion
with simultaneous spin-flip is a characteristic feature of the spin-orbit
coupling and means a virtual
conversion of the spin exciton into the cyclotron magnetoplasmon). As a
result, a new annihilation channel of the spin exciton scattering appears:
two spin excitons can be scattered by each other, within the same Landau level, directly due to
the hyperfine interaction -- finally one gets a single-spin exciton
state possessing the combined energy. This kind of scattering, as in the case
of scattering caused by disorder, \cite{di04,di09} is kinematic: the transition matrix element does not contain the Coulomb constant between bra- and ket-vectors -- the scattering is possible because the spin excitons  are not actually elementary Bose particles but possess an internal degree of freedom and thus have a ``memory'' of the Pauli principle for the primary  electron system.
Thus in spite of small hyperfine coupling constant, the hyperfine coupling channel competes with the spin-orbit ones and can even  dominate.

The next section of the paper is devoted to formal description of the
system where we present the Hamiltonian and the basis of exciton states
 (excitonic representation). In section III we study the hyperfine coupling relaxation mechanisms when the spin exciton annihilation/creation is determined by the spin-exciton - spin-exciton scattering including the dynamic and kinematic scattering channels. For this process  the relaxation rate is proportional to the spin-exciton number squared, and therefore the relaxation is non-exponential with time. (In principle, it becomes exponential when the spin exciton number approaches its equilibrium value, but the final exponential stage cannot, in fact,  be observed under  the condition $T\ll \epsilon_{\rm Z}$.) We discuss also in Sec. III possible relaxation processes, related to the hyperfine coupling and phonon emission/absorption, comparing them with other relaxation mechanisms. Section IV is devoted to the spin-orbit relaxation channels relevant to the considered region of strong magnetic fields and low temperatures. These spin-orbit mechanisms are also related to the spin-exciton - spin-exciton scattering but determined by two different dissipation processes -- via coupling to a smooth random potential or by coupling to phonons.

In Sec. V we discuss the results of our study. The main result  consists of the  interplay of different relaxation processes. We compare those due to the hyperfine coupling and spin-orbit interactions and, summing all relaxation channels, calculate the total characteristic inverse time. In this ``interplay regime'', where the spin-exciton - spin-exciton channels  dominate, the relaxation occurs non-exponentially, and the effective relaxation time
reaches its maximum $\sim 1-5\,\mu s$ depending on the Landau level fillings. Nevertheless, the relevant region of parameters $T$ and $B$ is not too extreme and experimentally quite accessible.

Note that in this paper we do not study the situation
where the Goldstone condensate of ``zero spin-excitons'' arises,\cite{di04,di11}, {\it i.e.} where there would be  a rotation of the {\it direction} but not a reduction in the  {\it magnitude} of the total spin of the system. Here we consider instead relaxation where there are, at low temperatures a bulk number of spin excitons arising from an intensive external (e.g., optical) excitation. The initial state at low temperatures should be described as a (metastable) ``thermodynamic condensate"  of  spin waves with non-zero, but small wave-vectors limited by the uncertainty determined by disorder.\cite{di04} We think that such  situation, where most of the spin exciton annihilation/creation events happen within the thermodynamic condensate, is  realisable experimentally.

\section{Formal Statement of the Problem: The Hamiltonian and the Basis of Exciton States}

Our system consists of two components: electrons belonging to the two-dimensional electron gas and nuclei of Ga and As atoms.  In
addition, we consider piezo- and  deformation couplings of the 2D electron gas
electrons to the lattice, which are reduced to electron-phonon
interaction. So, the Hamiltonian used is as follows:
\begin{equation}\label{tot_ham}
  {\hat H}={\hat H}_1+\epsilon_{\rm Z}{\hat S}_z+{\hat H}_{\rm int}+\sum_j{\hat H}_{\rm hf}^{(j)}+\sum_j{\hat
  H}^{(j)}_{\rm e-ph}\,.
\end{equation}
Here ${\hat H}_1=\sum_j\left[\hat{\bf q}_j^2/2m_e^*+H_{\rm so}^{(j)}\right]$ is single electron Hamiltonian including the spin-orbit coupling part ($\hat{{\bf q}}=\!-i{\bf\nabla}+e{\bf A}/c\hbar$); ${\hat S}_z=\sum_j{\hat \sigma}_z^{(j)}/2\;$; subscript $j$ labels electrons. The third term describes Coulomb energy of the {\em
e-e} interaction, the fourth is the hyperfine interaction of electrons with
nuclei, and the fifth is the operator of electron-phonon
interaction. If one holds $H_{\rm so}\!=\!0$, we can omit the orbital single electron energy terms --- all states relevant to our problem belong to the same Landau level,
and therefore have the same orbital energy equal to $\hbar\omega_c\nu
{\cal N}_{\phi}$ ($\omega_c$ is the cyclotron frequency, ${\cal N}_{\phi}$ is
the Landau level  degeneracy). We ignore also the energy of nuclei
which consists of the contribution due to their interaction independent of the
electrons, and of the nuclear Zeeman energy. Variations of both, associated with change of
nuclear spins, are negligibly small owing to the tiny nuclear magnetic momentum.

In the following three subsections (A, B, and C) we
neglect the spin-orbit coupling. The spin-orbit Hamiltonian and spin-orbit corrections, written   in terms of the representation used,  will be given in the subsection D.

\subsection{Electron system. Excitonic representation}
\label{A}

We now  present the basis set of states
diagonalizing the first three terms of the Hamiltonian
\ref{tot_ham} to leading order in parameter $r_{\rm
c}\!=\!(\alpha e^2/\kappa l_B)/\hbar\omega_c$ considered to be
small ($\alpha\!<\!1$ is the averaged form-factor arising  due to
finiteness of the 2D layer thickness; $\kappa$ is the dielectric
constant). We do this by analogy with previous
works,\cite{di04,di09,di96,2di96,di99} defining the spin exciton creation
operator$\,$\cite{dz83}
\begin{equation}\label{Q}
  {\cal Q}_{ab\,{\bf q}}^{\dag}=\frac{1}{\sqrt{ {\cal N}_{\phi}}}\sum_{p}\,
  e^{-iq_x p}
  b_{p+\frac{q_y}{2}}^{\dag}\,a_{p-\frac{q_y}{2}}\,,
\end{equation}
where $a_p$ and $b_p$ are the Fermi annihilation operators
corresponding to electron states on the upper Landau level with spin up
($a\!=\!\uparrow$) and spin down ($b\!=\!\downarrow$),
respectively. Index $p$ marks intrinsic Landau level states which have wave
functions $\psi_{n p}({\bf r})\!=\!(2\pi {\cal N}_\phi)^{-1/4}e^{ipy}\varphi_n(p\!+\!x)$ in the Landau
gauge. [$\varphi_n(x)$ is the oscillator function, where $n$ is number of the upper partially filled Landau level;  in Eq. (\ref{Q}) and everywhere below we
measure length in the $l_B$ units wave vectors in the $1/l_B$ ones.] In the {\it
odd-integer quantum Hall regime}, operator (\ref{Q}) acting on the
ground state yields the {\it eigenstate} of the first two terms
of Eq. (\ref{tot_ham}), namely:
\begin{equation}\label{eigen}
[\epsilon_{\rm Z}{\hat S}_z\!+\!{\hat H}_{\rm int},{\cal Q}_{ab\,{\bf
q}}^{\dag}]|0\rangle\!=\!(\epsilon_{\rm Z}\!+\!{\cal E}_q){\cal Q}_{ab\,{\bf
q}}^{\dag}|0\rangle,
\end{equation}
where $|{\rm
0}\rangle\!=\!|\overbrace{\uparrow,\uparrow,...\uparrow}^{\displaystyle{\vspace{-15mm}
\mbox{\tiny{$\;{\cal N}_\phi$}}}}\,\rangle$. This basic property of the
exciton state, ${\cal Q}_{ab\,{\bf q}}^{\dag}|0\rangle$, is
asymptotically exact to first order in $r_{\rm c}$. After
the introduction of intra-sublevel operators ${\cal A}^\dag_{\bf
q}\!=\!{\cal N}_{\phi}^{-1/2}{\cal Q}_{aa{\bf q}}^\dag$ and ${\cal
B}^\dag_{\bf q}\!=\!{\cal N}_{\phi}^{-1/2}{\cal Q}_{bb{\bf q}}^\dag$, we
obtain a closed Lie algebra for these exciton
operators.\cite{by87,di02,di07} The commutation identities needed
in our case are
\begin{equation}\label{commutators}
\begin{array}{l}
\displaystyle{\left[{\cal Q}_{{\bf q_1}},
  {\cal Q}_{{\bf q_2}}^{+}\right]=
  e^{i({\bf q}_1\times{\bf q}_2)_z/2}{\cal A}_{\bf q_1\!-
  \!q_2}-
  e^{-i({\bf q}_1\times{\bf q}_2)_z/2}{\cal B}_{\bf q_1\!-
  \!q_2},}\\
 \displaystyle{e^{-i({\bf q}_1\times{\bf q}_2)_z/2}[{\cal A}^\dag_{{\bf q}_1},
  {\cal Q}^\dag_{{\bf q}_2}]=
  -e^{i({\bf q}_1\times{\bf q}_2)_z/2}[{\cal B}^\dag_{{\bf q}_1},
  {\cal Q}^\dag_{{\bf q}_2}]=
  -{\cal N}_\phi^{-1} {\cal Q}^\dag_{{\bf q}_1\!+\!{\bf q}_2}}
\end{array}
\end{equation}
(Here and below we omit the subscript $ab$ at the ${\cal
Q}$-operators.) Note that the commutation algebra (\ref{commutators}) is neither
purely  Fermionic nor Bosonic.

The interaction Hamiltonian ${\hat H}_{\rm
int}=\frac{1}{2}\int\! d{\bf r}_1d{\bf r}_2\,{\hat \Psi}^\dag({\bf
r}_2){\hat \Psi}^\dag({\bf r}_1)U({\bf r}_1\!-\!{\bf r}_2){\hat \Psi}({\bf
r}_1){\hat \Psi}({\bf r}_2)$ may be
expressed in terms of the exciton operators.\cite{di02,di07} If we
keep in ${\hat H}_{\rm int}$ only the terms relevant to our
problem, it takes a very simple form
\begin{equation}\label{coul}
{\hat H}_{\rm int}=\frac{{\cal N}_\phi}{2} \sum_{\bf q}W({ q})
\left({\cal A}^\dag_{\bf q} {\cal A}_{\bf q} + 2 {\cal A}^\dag_{\bf
q}{\cal B}_{\bf q} + {\cal B}^\dag_{\bf q}{\cal B}_{\bf q}\right),
\end{equation}
Here $W({ q})=U(q)[f(q)]^2$, where $f=e^{-q^2/4}$ if $\nu\leq 1$, or
$f=e^{-q^2/4}[L_k(q^2/2)]$ if $\nu\!=\!2n\!+\!1$ ($L_n$ is the Laguerre polynomial). $U(q)$ is the Fourier component of the 2D
Coulomb
interaction function: $U(q)\!=\!(e^2/\kappa l_B q)\int\!\!\int
dz_1dz_2e^{-q|z_1-z_2|}|\chi(z_1)|^2|\chi(z_2)|^2$, where $\chi(z)$ is the dimensionally
quantized wave-function of an electron sized in the $z$-direction.

In contrast to integer quantum Hall ferromagnet, the use of the excitonic basis ${\cal
Q}_{{\bf q}}^{\dag}|0\rangle$ presents only a {\it model approach}
in the case of {\it fractional quantum Hall regime}. Generally, at
fractional filling, spin-flip excitations within the same Landau
level might have many-particle rather than two-particle nature
because the same change in the spin numbers (\ref{deltaS}) may be
achieved with participation of arbitrary number of
intra-spin-sublevel excitations (charge-density waves). These
waves are generated by the operator ${\cal A}^\dag_{\bf q}$ acting
on the ground state
 $|{\rm
0}\rangle\!=\!|\overbrace{\uparrow,..\uparrow,..\uparrow}^{\displaystyle{\vspace{-15mm}
\mbox{\tiny{$\;\nu {\cal N}_\phi$}}}}\,\rangle$.\cite{gi85} The result is
trivial in the case of integer $\nu$ (${\cal A}^\dag_{\bf q}|{\rm
0}\rangle\!=\!\delta_{{\bf q},\,0}|{\rm 0}\rangle$); however,
states of the ${\cal Q}_{{\bf q}_1}^{\dag}{\cal A}^\dag_{{\bf
q}_2}{\cal A}^\dag_{{\bf q}_3}...|0\rangle$ type might constitute
a basis set if one studies the spin-flip process at fractional
$\nu$. On the other hand, a comprehensive phenomenological
analysis$\,$\cite{gi85,lo93} suggests that even the spin-flip
basis reduced to single-mode (single-exciton) states would be
quite appropriate, at least for the lowest-energy excitations in
the case of fractional quantum Hall ferromagnet. This single-mode approach is
indirectly substantiated by the fact that the charge-density wave
has a Coulomb gap$\,$\cite{gi85} which is substantially  larger than the
Zeeman gap $\epsilon_{\rm Z}$. Hence for a fractional quantum Hall ferromagnet, just as
in Ref. \onlinecite{lo93}, we will consider the simple state ${\cal
Q}_{{\bf q}}^{\dag}|0\rangle$ to describe the spin-flip
excitation.  However, the calculation of $\langle 0|{\cal A}_{{\bf
q}}{\cal A}^\dag_{{\bf q}'}|0\rangle$ is required for the
following. Now this expectation is not simply equal to
$\delta_{{\bf q},\,0}\delta_{{\bf q}'\!,\,0}$, but is expressed in
terms of the two-particle correlation function $g(r)$ calculated
for the ground state:\vspace{-2mm}
\begin{equation}\label{CDWnorm}
  \langle 0|{\cal A}_{{\bf q}}{\cal A}^\dag_{{\bf
  q}'}|0\rangle=\displaystyle{\frac{\nu}{{\cal N}_\phi}}\left[2\pi\nu\overline{g}(q)e^{q^2/2}\!+\!1\right]\delta_{{\bf
  q}'\!,\,{\bf q}}\qquad (\nu\leq 1)\,.
\end{equation}
Here $\overline{g}(q)\!=\!\frac{1}{(2\pi)^2}\int\! g(r)e^{-i{\bf
qr}}d^2r$ is the Fourier component. The function $g(r)$ is known,
e.g., in the case of Laughlin's state.\cite{gi85,gi84} If the
ground state is described  by the Hartree-Fock approximation, we
have simply  $2\pi
\overline{g}\!=\!\left({\cal N}_\phi\delta_{q,\,0}\!-\!e^{-q^2/2}\right)$,
which does not depend on $\nu$. Furthermore, at odd-integer filling
factors this Hartree-Fock expression becomes a Fourier component
of the {\it exact} correlation function. In the latter case one
should also formally set $\nu\!=\!1$ in Eq. (\ref{CDWnorm}) Note
also that the exact equation $\langle 0|{\cal A}^\dag_{{\bf
q}}|0\rangle\!=\!\nu\,'\delta_{{\bf q},0}$ holds, where we set $\nu\,'\!=\!\nu$ if
$\nu\leq 1$ but $\nu\,'\!=\!1$ if $\nu\!=\!3,5,..$.

With the help of Eqs. (\ref{commutators}) and (\ref{coul}) one can check Eq. (\ref{eigen})
in the case of odd-integer $\nu$. If $\nu$ is fractional, the Coulomb exciton energy
within the single mode approximation is defined as
${\cal E}_q\!=\!\langle 0|{\cal Q}_{\bf q}[{\hat H}_{\rm int},{\cal Q}_{\bf q}^\dag]|0\rangle/\langle 0|{\cal Q}_{\bf q}{\cal Q}_{\bf q}^\dag|0\rangle$.\cite{lo93}
As a result
in both cases of integer or fractional
$\nu<1$ one obtains for small $q$ the quadratic dispersion law (\ref{spectrum}) with the  spin exciton mass
\begin{equation}\label{mass}
1/M_{\rm x}=\frac{1}{2}\int_0^\infty W(q)q^3\left(1-\frac{{\cal N}_\phi}{\nu\,'}\langle 0|{\cal A}_{{\bf q}}{\cal A}^\dag_{{\bf
  q}}|0\rangle\right)dq\,.
\end{equation}
We have employed  the rule for change from summation to integration over the 2D vector ${\bf q}$: $\sum_{\bf q}...=\displaystyle{{\cal N}_\phi\!\!\int\!...\,qdqd\phi/2\pi}$.

\subsection{Nuclear system. Hyperfine coupling  in  the excitonic representation. Hybridization of
spin-exciton and nuclear spin-flip states.}

The general expression of the hyperfine coupling Hamiltonian$\,$\cite{la77} is simplified in the case of interaction with nuclei in a semiconductor matrix.\cite{pa77,dy84,me02} As this simplification
is valid in the 2D channel of a quantum well we may directly start from the well known expression for contact interactions of electrons with nuclei
\begin{equation}\label{HF}
 {\hat H}_{\rm hf}=\frac{v_0}{2}\sum_nA_n\Psi^*(\mbox{\boldmath
 $R$}_n)\left(\hat{\mbox{\boldmath $I$}}^{(n)}\!\!\!\!\cdot\!\hat{\mbox{\boldmath
 $\sigma$}}\right)\Psi(\mbox{\boldmath $R$}_n)\,,
\end{equation}
(see for example Ref. \onlinecite{dy84} and
references therein.)
where $\hat{\mbox{\boldmath $I$}}^{(n)}$ and $\mbox{\boldmath $R$}_n$ are spin and position of
the $n$-th nucleus and $\Psi(\mbox{\boldmath $R$})$ is the envelope function of electron
[$\mbox{\boldmath $R$}=({\bf r},z)$ is the 3D vector]. Both Ga and As nuclei have the same
total spin: $I^{\rm Ga}\!=\!I^{\rm As}\!=\!3/2$. In Eq. (\ref{HF}) $v_0$ is volume of the unit
cell. The parameter $A_n$, being inversely proportional to $v_0$, really depends only on position of the Ga/As nucleus within the unit cell. For the final calculation we need the sum $A_{\rm Ga}^2\!+\!A_{\rm
As}^2$. If $v_0$ is volume of the two atom
unit cell then $A_{\rm Ga}^2\!+\!A_{\rm
As}^2\!\simeq\!4\cdot 10^{-3}\,$meV${}^2$. (See Appendix A.)

Now we rewrite $\hat{\mbox{\boldmath $I$}}^{(n)}\!\!\!\!\!\cdot\!\hat{\mbox{\boldmath
$\sigma$}}$ as ${\hat I}_z{\hat \sigma}_z\!+\!{\hat I}_+{\hat \sigma}_-\!+\!{\hat I}_-{\hat
\sigma}_+$. Then omitting the ${\hat I}_z{\hat \sigma}_z$ term due to its irrelevance to any
spin-flip process and substituting in Eq. (\ref{HF}) the Schr\"odinger operators ${\hat
\Psi}^\dag(\mbox{\boldmath $R$})\!=\!\chi(z)\sum_p\left(a^\dag_p\!+\!b^\dag_p\right)\psi_p^*(
{\bf r})$ and
${\hat \Psi}(\mbox{\boldmath $R$})\!=\!\left({\hat \Psi}^\dag\right)^\dag$  instead of ${
\Psi}^*$ and ${ \Psi}$, we come to
\begin{equation}\label{HF2}
{\hat H}_{\rm
hf}=\frac{v_0}{2}\sum_{p_1,p_2}b^\dag_{p_2}a_{p_1}\sum_n|\chi(Z_n)|^2\psi_{p_2}^*(X_n,Y_n)\psi_{p_1}(X_n,Y_n)A_n{\hat
I}_+^{(n)}+\quad\mbox{H.c.}
\end{equation}

Substitution of the equation
\begin{equation}\label{ba-inverse}
  b^\dag_{p_2}a_{p_1}=\sum_{\bf
  q}\frac{e^{iq_x(p_2\!-\!q_y/2)}}{\sqrt{{\cal N}_\phi}}\delta_{q_y,p_2\!-\!p_1}{\cal Q}^\dag_{\bf
  q}\,,
\end{equation}
which is simply inverse to Eq. (\ref{Q}), yields after summation over $p_1$ and $p_2$ the hyperfine coupling
Hamiltonian in the excitonic representation:
\begin{equation}\label{HF3}
{\hat H}_{\rm hf}=\frac{v_0}{4\pi l_B^2\sqrt{{\cal N}_\phi}}\sum_{\bf q}f(q){\cal Q}_{\bf
q}\sum_nA_n|\chi(Z_n)|^2e^{i{\bf q}{\scriptsize\mbox{\boldmath $R$}}_n}{\hat
I}^{(n)}_-\quad+\quad\mbox{H.c.}
\end{equation}

A set of the $I_z$ spin numbers $\{M\}\!=\!(M_1,M_2,...M_n,...)$, where every $M_n$ may take
on values
$-3/2,-1/2,1/2,3/2$, completely determines the state of the nuclear system. The
state where 2D electrons are in the ground state and nuclei in the state $\{M\}$ we denote
as $\left|\{M\},0\right\rangle$. By applying  the lowering/raising
operator $I^{(n)}_\mp$ to this state, we obtain
\begin{equation}\label{action}
 {\hat I}^{(n)}_\mp\left|\{M\},0\right\rangle=\sqrt{\left(\frac{5}{2}\!\mp\!
 M_n\right)\left(\frac{3}{2}\!\pm\! M_n\right)}\left|\{M\}^\mp_n,0\right\rangle,
\end{equation}
where $\{M\}^\mp_n=(M_1,M_2,...M_n\!\mp\!1,...)$. Let us find the hyperfine coupling correction to the
normalized spin exciton state ${\cal Q}^\dag|\{M\},0\rangle/\sqrt{\nu\,'}$.  Considering operator (\ref{HF3}) as a
perturbation we obtain with the help of Eqs. (\ref{commutators}) and (\ref{action}):
\begin{equation}\label{SE}
\begin{array}{l}
\displaystyle{|{\rm SE},{\bf q}\rangle={{\cal Q}^\dag_{\bf
q}|\{M\},0\rangle}/{\sqrt{\nu\,'}}}\\
\qquad\quad\displaystyle{+\frac{v_0\sqrt{\nu\,'}f(q)}
{4\pi\sqrt{{\cal N}_\phi}l_B^2E_{\rm x}(q)}\sum_nA_n|\chi(Z_n)|^2e^{i{\bf
q}{\mbox{\scriptsize\boldmath $R$}}_n}\sqrt{\left(\frac{5}{2}\!-\!
M_n\right)\left(\frac{3}{2}\!+\! M_n\right)}\left|\{M\}^-_n,0\right\rangle}\,.
\end{array}
\end{equation}
In the same way we find the corrected nuclear `spin-turned' (NST) state
\begin{equation}\label{NST}
\begin{array}{l}
\displaystyle{|{\rm NST},n\rangle=\left|\{M\}^-_n,0\right\rangle}\\
-\displaystyle{\frac{v_0}
{4\pi\sqrt{{\cal N}_\phi}l_B^2}\sum_{{\bf q},n'}\frac{f(q)}{E_{\rm
x}(q)}A_{n'}|\chi(Z_{n'})|^2e\!{}^{-i{\bf q}{\mbox{\scriptsize\boldmath
$R$}}_{n'}}\sqrt{\left(\frac{5}{2}\!+\! M_{n'}\!-\!\delta_{n',n}\right)\left(\frac{3}{2}\!-\!
M_{n'}\!+\!\delta_{n',n}\right)}{\cal Q}_{\bf q}^\dag\left|\{M\}^{-+}_{nn'},0\right\rangle},
\end{array}
\end{equation}
where we consider $M_n>-3/2$, and use notation
$\{M\}_{nn'}^{-+}=(M_1,...M_n\!-\!1,...M_{n'}\!+\!1,...)$ meaning by that $\{M\}_{nn}^{-+}\equiv
\{M\}$.
[The $n'\!=\!n$ term in the sum of Eq. (\ref{NST}) contributes to the transition matrix
element relevant to some spin exciton relaxation processes.]

The hybridized states (\ref{SE}) and (\ref{NST}) diagonalize the first three terms of the
Hamiltonian (\ref{tot_ham}) to the first order in hyperfine coupling. Correspondingly, these have energies
$E_{\rm x}(q)$ and $0$ (counted from the energy of the $|\{M\},0\rangle$ state) within the
approximation neglecting energy corrections of the second order in hyperfine coupling and small magnetic
energy corrections related to changes of $M_n$ momenta.

\subsection{Electron-phonon interaction in the exciton representation.}

The Hamiltonian of the interaction of electrons with 3D acoustic phonons is written
as:\cite{io89}
\begin{equation}\label{e-ph1}
{\hat H}_{\rm e-ph}=\frac{{\hbar}^{1/2}}{L L_z^{1/2}}
  \,\sum_{{\bf q},{k}_z,s}
  {U'}_{s}({\bf k})\,
  {\hat P}_{{\bf k},s}
{\cal H}_{{\rm e-ph}}({\bf q})\qquad+\qquad
  \mbox{H.~c.}\,,
\end{equation}
where $L^2\!=\!2\pi {\cal N}_\phi l_B^2$ is the 2D area, and $L_z$ is the dimension of the sample
along ${\hat z}$,
\begin{equation}\label{e-ph2}
{\cal H}_{\rm e-ph}({\bf q})=
   \int e^{i{\bf qr}}
   {\hat \Psi}^{\dag}({\bf r}){\hat \Psi}({\bf r})\,d^{2}r\,,\:\quad
   {\bf k}= ({\bf q},k_z)\,;
\end{equation}
${\hat P}_{{\bf k},s}$ is the phonon annihilation operator (index $s$ denotes
possible phonon polarizations: the longitudinal  $l$ or one  of the two transverse polarizations $t_1$ or $t_2$), and
${U'}_{s}({\bf k})$ is the renormalized
vertex which includes the fields of deformation (DA) and piezoelectric
(PA) interactions. The integration with respect to $z$ has been already
performed and leads to the renormalization
${ U'}_{s}({\bf k})=
U_{s}({\bf k})\int \chi^{*}(z)e^{ik_{z}z}\chi(z)\,dz$

The isotropic model for the phonon field${\,}$\cite{gale87} enables us to take into account
the deformation and piezoelectric couplings independently.  We further use the approximation where
we take no difference between longitudinal and transverse sound velocities. For the
three-dimensional (3D) vertex one needs only the expressions for the
squares,\cite{io89,gale87}
\begin{equation}\label{vertex}
  |U_s|^2=\pi\varepsilon_{\rm ph}({\bf k})/p_0^3\tau_{s}({\bf k})\,,
\end{equation}
where the phonon energy is $\varepsilon_{\rm ph}=\hbar c\sqrt{k_z^2\!+\!q^2}/l_B$ (we recall that $k_z$ and ${\bf q}$ are dimensionless), $p_0=2.52\cdot
10^6\,$cm${}^{-1}$ is the material parameter of GaAs (see Ref.
\onlinecite{gale87}). The longitudinal $\tau_{ l}({\bf k})$ and transverse
$\tau_{t}({\bf k})$ times are the 3D acoustic
phonon life-times (see Appendix~A). These quantities
are expressed in terms of nominal times $\tau_D$ and
$\tau_P$ characterizing respectively DA and PA phonon
scattering in three-dimensional GaAs crystal.
(See Appendix~A, Ref. \onlinecite{2di96} and cf. Ref. \onlinecite{gale87}.)

The dimensionless operator
${\cal H}_{\rm e-ph}$
in terms of the excitonic representation has the following simple form (cf. Ref. \onlinecite{di00})

\begin{equation}\label{e-ph3}
{\cal H}_{\rm e-ph}({\bf q})=f(q){\cal N}_\phi\left({\cal A}_{{\bf q}} +
  {\cal B}_{{\bf q}}\right).
\end{equation}

\subsection{The spin-orbit coupling in the excitonic representation.}
\label{D}

If considering the spin-orbit coupling, we will ignore the hyperfine coupling but take into account the ${H}_{\rm so}$ operator in the single electron part ${\hat H}_1$ of the
Hamiltonian \eqref{tot_ham}:
\begin{equation}\label{SO}
  {\hat H}_{\rm so}=\alpha\left(\hat{{\bf q}}\times\hat{\mbox{\boldmath $\sigma$}}
  \right)_{\!z}\!+\!
  \beta\left(\vphantom{\left(\hat{{\bf q}}\times\hat{\mbox{\boldmath
  $\sigma$}}
  \right)}\hat{ q}_y\hat{\sigma}_y\!-\!{\hat q}_x\hat{\sigma}_x\right),\qquad \hat{{\bf q}}=-i{\bf\nabla}+e{\bf A}/c\hbar\,.
\end{equation}
This operator, specified for the
(001) GaAs plane, represents a combination of the Rashba term ($\sim\alpha$) and the crystalline anisotropy term ($\sim\beta$)\,\cite{by84} and does not violate  translational symmetry.\cite{foot1}

Now it is convenient to use a bare single-electron basis diagonalizing Hamiltonian $\hat{\bf q}^2/2m_e^*+H_{\rm so}$.
To within the leading order in the $H_{\rm so}$ terms we obtain
\begin{equation}\label{so_ab}
\Psi_{pa}=
  \left( {\psi_{n p}\atop v\sqrt{n\!+\!1}\psi_{n\!+\!1\,p}
  +iu\sqrt{n}\psi_{n\!-\!1\,p}}\right)\:\;\mbox{and}\:\;
  \Psi_{pb}=\left({-v\sqrt{n}
  \psi_{n\!-\!1\,p}
  +iu\sqrt{n\!+\!1}\psi_{n\!+\!1\,p}\atop
  \psi_{n p}}\right),
\vspace{-1mm}
\end{equation}
where $u$ and $v$ are small dimensionless parameters:
$u=\beta\sqrt{2}/\l_B\hbar\omega_c$ and $v=\alpha\sqrt{2}/\l_B\hbar\omega_c$. Thus the single-electron states acquire a chirality $a$ or $b$ instead of spin quantum number, and the spin flip corresponds
to the $a\to b$ process. The definition of the spin exciton creation operator formally remains the same [Eq. \eqref{Q}], however the $a_p$ and $b_p$ operators describe annihilation in the states \eqref{so_ab} now.

When being presented in terms
of basis states \eqref{so_ab}, spin operators $\int \Psi^{\dag}{\hat {\bf S}}^2\Psi d^2{\bf r}$ and $\int
\Psi^{\dag}{\hat { S}}_z\Psi d^2{\bf r}$ [where
$\Psi\!=\!\sum_p(a_p\Psi_{pa}\!+\!b_p\Psi_{pb})$] are invariant  up to the second order of
$u$ and $v$. However, in the excitonic representation the interaction Hamiltonian ${\hat H}_{\rm int}$ and the electron-phonon coupling operator acquire terms proportional to $u$ and $v$, which are additional to Eqs. \eqref{coul} and \eqref{e-ph3} respectively.\cite{di96,2di96,di99,di00} These terms correspond to creation and annihilation of spin excitons in the system:
\begin{equation}\label{so_coul}
{\hat H}'_{\rm int}=
  {\cal N}_{\phi}^{1/2}\sum_{\bf q}(iuq_+-vq_-)W(q)\left({\cal A}_{\bf q}^\dag
  +{\cal B}_{\bf q}^\dag\right){\cal Q}_{\bf q} + \mbox{H. c.},
\end{equation}
and
\begin{equation}\label{so_e-ph3}
{\cal H}_{\rm e-ph}'({\bf q})={\cal N}_\phi^{1/2}f(q)
\left(iuq_+-vq_-\right)
  {\cal Q}_{\bf q} + \mbox{H. c.}
\end{equation}
$\left[q^\pm\!=\!\mp i(q_x\!\pm\!iq_y)\!/\!\sqrt{2}\right]$.

We can also take into account the presence of an external smooth random potential $\varphi({\bf r})$. This  is assumed to be Gaussian and  defined by a correlator $K({\bf
r})=\langle \varphi({\bf r})\varphi(0)\rangle$. By choosing $\langle \varphi({\bf r})\rangle=0$, , the correlator is
$  K({\bf r})=\Delta^2\exp{(-r^2/\Lambda^2)}$, in terms of the correlation length $\Lambda$ and the amplitude $\Delta$. The smooth random potential can act as the rate-limiting process in the energy dissipation which makes the spin-flip process irreversible.  $\varphi({\bf r})$
formally is analogous to frozen field of phonons having zero frequency. A static potential  cannot cause dissipation alone: physically the random potential (mixed with the spin-orbit term) causes spin-flip and breaks momentum conservation. The actual dissipation comes from other interactions that do not change the spin: electron-electron and electron-phonon interactions that occur on a faster time scale and render the process irreversible. Therefore, using again the Eq. \eqref{so_ab} basis set and Eq. \eqref{ba-inverse}, we obtain the $\hat{\varphi}$ operator in terms of the excitonic representation. The part responsible for a spin-flip is\,\cite{di04,di09}
\begin{equation}\label{srp}
 \hat{\varphi}'={\cal N}_{\phi}^{1/2}\sum_{\bf q}f(q)
  \overline{\varphi}
  ({\bf q})\left(iuq_+-vq_-\right)
  {\cal Q}_{\bf q}+\mbox{H.c.}\,,
\end{equation}
where $\overline{\varphi}$ is the Fourier component [$\varphi({\bf r})\!=\!\sum_{\bf
q}\!\overline{\varphi}({\bf q})e^{i{\bf qr}}$].

\section{The spin-exciton - spin-exciton scattering relaxation channels governed by the hyperfine coupling}

The $\delta S_z\!=\!-1$ hybridized states (\ref{SE}) and (\ref{NST}) diagonalize the Hamiltonian ${\hat H}_{\rm int}\!+{\hat H}_{\rm hf}$, but the $\delta S_z\!=\!-2$ states ${\cal Q}_{{\bf q}_1}^\dag|{\rm SE},{\bf q}_2\rangle$ and ${\cal Q}_{{\bf q}}^\dag|{\rm NST},n\rangle$ do not. (Here by  $S_z$ we mean  the total spin
number of the combined nuclear and electron system.) The problem may be
formulated in terms of a scattering where the double exciton state ${\cal Q}_{{\bf q}_1}^\dag|{\rm SE},{\bf q}_2\rangle$ transforms to the single exciton one ${\cal Q}_{{\bf q}}^\dag|{\rm NST},n\rangle$. Since the hyperfine coupling energy is neglected, the energy conservation law takes the form
\begin{equation}\label{conservation}
E_{\rm x}({ q}_1)+E_{\rm x}({ q}_2)=E_{\rm x}({ q})\,.
\end{equation}
It determines the modulus of the spin exciton momentum ${\bf q}$ in the final state, and in particular means that ${\bf q}$ cannot be equal to ${\bf q}_1$ or ${\bf q}_2$.

\subsection{Kinematic scattering}

The transition matrix element ${\cal M}_{if}$ in Eq. (\ref{FGR}) has to be found to  first order in the hyperfine coupling.  Therefore in the case of the kinematic scattering, where ${\cal M}_{if}$ represents an expectation value $\langle{\rm bra}|{\hat H}_{\rm hf}|{\rm ket}\rangle$  calculated directly for the hyperfine coupling operator, the ket- and bra-vectors are determined only by the main components of the ${\cal Q}_{{\bf q}_1}^\dag|{\rm SE},{\bf q}_2\rangle$ and ${\cal Q}_{{\bf q}}^\dag|{\rm NST},n\rangle$ states without any hyperfine coupling corrections. Namely, taking into account that the initial double-exciton state and the final single-exciton one have to be normalized, we should calculate the kinematic scattering matrix element
\begin{equation}\label{M_kin}
{\cal M}_{if}^{\rm kin}({\bf q}_1,{\bf q}_2,{\bf q},n)=\langle 0,\{M\}_n^-|{\cal Q}_{{\bf q}}{\hat H}_{\rm hf}{\cal Q}_{{\bf q}_1}^\dag{\cal Q}_{{\bf q}_2}^\dag|\{M\},0\rangle/{\nu\,'}^{3/2}\,.
\end{equation}
After substitution of Eqs. \eqref{HF3} and \eqref{action}, this is reduced to calculation of the four-operator expectation value (C.1) (see Appendix C). Note that were the ${\cal Q}$-operators usual Bose operators, the expectation (C.1) would simply be equal to $\delta_{{\bf q}'\!,\,{\bf q}_1}\delta_{{\bf q},\,{\bf q}_2}\!+\delta_{{\bf q}'\!,\,{\bf q}_2}\delta_{{\bf q},\,{\bf q}_1}$,  the conservation condition (\ref{conservation}) could not be satisfied and therefore the kinematic scattering channel would not  exist. Therefore {\it only due to the non-Bose nature of the spin-exciton states does this relaxation mechanism take place.}

We should keep in ${\cal M}^{\rm kin}_{if}$ only the main terms contributing to the final result, namely to the relaxation rate calculated on the basis of the Fermi golden rule (\ref{FGR}) and subsequent  summation over the ${\bf q}_1$, ${\bf q}_2$ and ${\bf q}$ statistical distributions. These are terms to the lowest power of ${ q}_1$ ${ q}_2$ and ${ q}$. They give the exact result to  leading order in the small parameter $TM_{\rm x}$. ($T$ is the temperature, characteristic values of
the momenta are $q_1,q_2,q\sim\sqrt{TM_{\rm x}}\ll 1$.) In particular, one  finds
that the $\sim \nu/{\cal N}_\phi$ terms in Eq. (C.1) give the strongest  contribution, and the $\sim \langle 0|{\cal A}_{...}{\cal A}_{...}^\dag|0\rangle$ terms may be neglected.\cite{g-function}
In addition, the terms where ${\bf q}\!=\!{\bf q}_1$ or ${\bf q}\!=\!{\bf q}_2$, are omitted due to the `selection rule' determined by Eq. (\ref{conservation}). As a result we obtain
\begin{equation}\label{M_kin2}
{\cal M}_{if}^{\rm kin}({\bf q}_1,{\bf q}_2,{\bf q},n)=
=-\displaystyle{\frac{v_0A_n|\chi(Z_n)|^2}{2\pi l_B^2}\sqrt{\frac{\left(\frac{3}{2}\!+\! M_{n}\right)\left(\frac{5}{2}\!-\!
M_n\right)}{{{\cal N}_\phi^3\nu\,'}}}\,e{}^{i({\bf q}_1\!+\!{\bf q}_2\!-\!{\bf q}){\mbox{\scriptsize\boldmath
$R$}}_{n}}}\,.
\end{equation}

\subsection{Dynamic scattering}

 If studying the dynamic scattering, one should take into account that the Coulomb interaction operator (\ref{coul}), acting on a certain state, does not change the number of the spin exciton operators determining this state, -- i.e. this number must be the same in the bra- and ket-states contributing to
 ${\cal M}_{if}\!=\!\langle{\rm bra}|{\hat H}_{\rm int}|{\rm ket} \rangle$. Furthermore the Coulomb interaction does not change the   total momentum of the electron gas, -- it too must be the same in the bra- and ket-states. Therefore, again only  the $\sim {\cal Q}_{{\bf q}_1}^\dag{\cal Q}_{{\bf q}_2}^\dag|\{M\},0\rangle$ component should be kept in the initial state ${\cal Q}_{{\bf q}_1}^\dag|{\rm SE},{\bf q}_2\rangle$. [The hyperfine coupling correction component can contribute only to the transition where ${\bf q}\!=\!{\bf q}_1$, which is forbidden due to Eq. (\ref{conservation})].

The single exciton state ${\cal Q}_{\bf q}^\dag|0\rangle$ diagonalizes the Hamiltonian ${\hat H}_{\rm int}$, but the double exciton state ${\cal
Q}_{{\bf q}_1}^\dag{\cal Q}_{{\bf q}_2}^\dag|0\rangle$ does not. The latter is in fact an `almost'
eigenstate. Indeed, even at odd-integer $\nu$ we have
[cf. Eq. (\ref{eigen})]
\begin{equation}\label{eigen2}
[{\hat H}_{\rm int},{\cal Q}_{{\bf
q}_1}^{\dag}{\cal Q}_{{\bf
q}_2}^{\dag}]|0\rangle\!=\!\left({\cal E}_{q_1}\!+
{\cal E}_{q_2}\!\right){\cal Q}_{{\bf q}_1}^{\dag}{\cal Q}_{{\bf q}_2}^{\dag}|0\rangle+
\left[[{\hat H}_{\rm int},{\cal Q}_{{\bf q}_1}^{\dag}],
{\cal Q}_{{\bf q}_2}^{\dag}\right]|0\rangle,
\end{equation}
where the double-commutation term arises due to the interaction between the
spin excitons. It can be routinely calculated with the help of Eqs. (\ref{commutators}) and (\ref{coul}), see Eq. (C.2) in Appendix C.  The norm of this term and the
averaged spin-exciton - spin-exciton interaction energy $\langle 0|{\cal Q}_{{\bf q}_2}{\cal
Q}_{{\bf q}_1}|\left[[{\hat H}_{\rm int},{\cal Q}_{{\bf q}_1}^{\dag}],{\cal
Q}_{{\bf q}_2}^{\dag}\right]|0\rangle$, both vanishing if $q_1\!=\!0$ or $q_2\!=\!0$, are respectively
$\lapprox(\alpha e^2/\kappa l_B){\cal N}_\phi^{-1/2}$ and $\lapprox(\alpha e^2/\kappa l_B)/{\cal N}_\phi$ if
$ q_1q_2\!\not=\!0$. The latter estimation quite
corresponds to an effective mean dipole-dipole interaction of two spin excitons sized within the area $2\pi l_B^2{\cal N}_\phi$. [We recall  that each magneto-exciton possesses a dipole momentum equal to $e({\bf q}\times {\hat z})l_B^2$ (in
common units).\cite{go68}]

It follows from the above that for the dynamic scattering process we choose the ket- and bra-vectors as
\begin{equation}\label{ini-fin}
\begin{array}{l}
\!\!|{\rm ket}\rangle={\cal Q}_{{\bf q}_1}^\dag{\cal Q}_{{\bf q}_2}^\dag|\{M\},0\rangle/\nu\,',\\
\!\!\mbox{and}\quad\langle{\rm bra}|=\langle n,{\bf q}|\\
\!\!=-\displaystyle{\frac{v_0}
{4\pi\sqrt{\nu\,'{\cal N}_\phi}l_B^2}A_{n}|\chi(Z_{n})|^2\sqrt{\left(\frac{3}{2}\!+\! M_{n}\right)\left(\frac{5}{2}\!-\!
M_n\right)}\left\langle0,\{M\}\right|}{\cal Q}_{{\bf q}}\sum_{{\bf q}'}{\cal Q}_{{\bf q}'}\frac{f(q')}{E_{\rm
x}(q')}e\!{}^{i{\bf q}'{\mbox{\scriptsize\boldmath
$R$}}_{n}}\,,
\end{array}
\end{equation}
implying that only the hyperfine coupling correction term is relevant in the final normalized state ${\cal Q}_{{\bf q}}^\dag|{\rm NST},n \rangle/{\nu\,'}^{1/2}$.
The matrix element meant to be calculated is
\begin{equation}\label{M_dyn1}
{\cal M}_{if}^{\rm dyn}({\bf q}_1,{\bf q}_2,{\bf q},n)=\left\langle n, {\bf q}\left|\left[[{\hat H}_{\rm int},{\cal Q}_{{\bf
q}_1}^{\dag}],{\cal Q}_{{\bf
q}_2}^{\dag}\right]\right|\{M\},0\right\rangle/\nu\,'\,.
\end{equation}
By analogy with the kinematic scattering, we keep in ${\cal M}^{\rm dyn}_{if}$ only terms to  the lowest power of ${ q}_1$ ${ q}_2$ and ${ q}$. Using sequentially Eqs. (C.2), (C.1), (\ref{mass}) and (3.1), we  find
\begin{equation}\label{M_dyn2}
{\cal M}_{if}^{\rm dyn}({\bf q}_1,{\bf q}_2,{\bf q},n)=
-\frac{{\bf q}_1{\bf q}_2\,{\cal M}_{if}^{\rm kin}({\bf q}_1,{\bf q}_2,{\bf q},n)}{q^2+{\bf q}_1{\bf q}_2-{\bf q}({\bf q}_1\!+{\bf q}_2)}\,.
\end{equation}

\subsection{The relaxation rate}

To calculate the spin-wave relaxation rate, one should know the distribution $N_{\bf q}$ of spin excitons over the ${\bf q}$ wave numbers. Although exciton operators \eqref{Q} are non-bosonic, the  spin exciton obey Bose statistics, because their number in any state determined by a certain ${\bf q}$ may, in principle, be macroscopically large. At any moment the spin excitons distribution is in quasi-equilibrium and characterized by a chemical potential $\mu\!<\!\epsilon_{\bf Z}$. (The thermodynamic equilibrium is established much faster than spin-flip processes occur.) Initially the total number
 of spin excitons $\,{\cal N}_{\bf x}\!=\!\nu\,' {\cal N}_\phi/2\!-\!S\,$  is actually determined
 by a short external optical impulse, and its value might be even more than the critical value
 \begin{equation}\label{N_c}
 {\cal N}_{{\bf x}c}\!=\!{\cal N}_\phi\int_{q_0}^\infty\displaystyle{\frac{qdq}{\exp{({\cal E}_q/T)}\!-\!1}}\!=\!{\cal N}_\phi M_{\rm x}T\left[q_0^2/2M_{\rm x}T-\ln{\left(e^{q_0^2/2M_{\rm x}T}\!-\!1\right)}\right]\,,
 \end{equation}
where we have used the quadratic approximation \eqref{spectrum} and designated as $q_0$ a lowest limit of possible nonzero values of $q$. Any violation of the translation symmetry contributes to the estimation of $q_0$.
For example in the ideally clean case $q_0\sim 1/L$, where $L\sim \sqrt{{\cal N}_\phi}$ is the linear dimension of the 2D system. A more realistic estimation can be made  if one takes into account the presence of a smooth random potential, then $q_0\!\sim\!M_{\rm x}l_B\Delta/\Lambda$, where $\Delta$ is the potential amplitude ($\Delta\!\ll\!1/M_{\rm x}$), and $\Lambda$ is the correlation length ($\Lambda\!\gg\!l_B$).\cite{di04} In practice $q_0\lapprox 0.01$. If ${\cal N}_{\rm x}\!>\!{\cal N}_{{\rm x}c}$, then the bulk number of spin excitons with nonzero but momenta $|{\bf q}|\lapprox q_0$ form a thermodynamic condensate. The specific $q$-distribution of excitons within the condensate plays no role; however we may write
\begin{equation}\label{distr1}
    N_{\bf q}=\left\{
 N_{\bf q}^{\{0\}} \,,\qquad\qquad\quad\:\;\, {\mbox{if}}\quad {\bf q}\in \{0\} \atop
  \displaystyle{1/\left[{\exp{({\cal E}_q/T})\!-\!1}\right]}\,,\:\;\;\;\! {\mbox{if}}\quad {\bf q}\notin\{0\}\quad\right.\,
\end{equation}
(${\bf q}\in \{0\}$ means belonging to the thermodynamic condensate). The number of the condensate excitons is thus ${\cal N}_{\rm x}\!-\!{\cal N}_{{\rm x}c}\!=\!\sum\limits_{{\bf q}\in \{0\}}\!\!N_{\bf q}^{\{0\}}$. During the spin exciton relaxation process the condensate is depleted, and when ${\cal N}_{\rm x}\!<\!{\cal N}_{{\rm x}c}$ we have:
\begin{equation}\label{distr2}
N_{\bf q}=1/\left[{\exp{({\cal E}_q\!+\!\epsilon_{\rm Z}\!-\!\mu)/T}\!-\!1}\right]\,,
\end{equation}
with chemical potential equal to
\begin{equation}\label{mu}
\mu=\epsilon_{\rm Z}+T\ln{\left[1-\exp\left(-\frac{{\cal N}_{\rm x}}{{\cal N}_\phi M_{\rm x}T}\right)\right]}\,.
\end{equation}
[In the vicinity of $\epsilon_{\rm Z}$ the value $\mu$ is determined with an accuracy: $|\mu-\epsilon_{\rm Z}|\gapprox \min(q_0^2/2M_{\rm x},\,T)$.] The $\mu=0$ equation determines the equilibrium spin exciton number: ${\cal N}_{\rm x}^{(0)}=-{\cal N}_\phi M_{\rm x}T\ln{\left(1-e^{-\epsilon_{\rm Z}/T}\right)}$.

The spin wave relaxation rate is defined as the difference between the fluxes of annihilating and created spin excitons in the phase space:
\begin{equation}\label{rate}
-\frac{d{\cal N}_{\bf x}}{dt}\!=\!\frac{1}{2}\sum_{{\bf q}_{1},{\bf q}_2}S({\bf q}_1,{\bf q}_2)
  \left[N_{{\bf q}_1}N_{{\bf q}_2}\left(1+N_{12}\right)-N_{12}\left(1+N_{{\bf q}_1}\right)
  \left(1+N_{{\bf q}_2}\right)\right],
\end{equation}
where $N_{12}=1/\left[{e^{({\cal E}_{q_1}\!+{\cal E}_{q_2}\!+\epsilon_{\rm Z})/T}\!-\!1}\right]$ if ${\cal N}_{\rm x}\!>\!{\cal N}_{{\rm x}c}$ or $N_{12}=1/\left[{e^{({\cal E}_{q_1}\!+{\cal E}_{q_2}\!+2\epsilon_{\rm Z}\!-\!\mu)/T}\!-\!1}\right]$ if ${\cal N}_{\rm x}\!<\!{\cal N}_{{\rm x}c}$,
and
the summation over final state values ${\bf q}$ is performed by calculating
\begin{equation}\label{Msquared}
\begin{array}{r}
\!\!\!\!\!S({\bf q}_1,{\bf q}_2)=\displaystyle{\frac{2\pi}{\hbar}
\sum_n\sum_{{\bf q}}\left|{\cal M}^{\rm kin}_{if}({\bf q}_1,{\bf q}_2,{\bf q},n)+{\cal M}^{\rm dyn}_{if}({\bf q}_1,{\bf q}_2,{\bf q},n)\right|^2\!\delta({\cal E}_{q}\!-{\cal E}_{{ q}_1}\!-{\cal E}_{{ q}_2}\!-\epsilon_{\rm Z})}\vspace{0mm}\\\qquad\displaystyle{=\frac{v_0^2M_{\rm x}\left[1+{\cal F}({q}_1^2,{q}_2^2,\phi,2M_{\rm x}\epsilon_{\rm Z})\right]}{2\pi {\cal N}_\phi^2\hbar\nu\,'l_B^4}
\sum_n\left(\frac{3}{2}\!+\!M_n\right)\left(\frac{5}{2}\!-\!M_n\right)A_n^2
|\chi(Z_n)|^4}\\\qquad\qquad\qquad\qquad\displaystyle={{\left[1+{\cal F}({ q}_1^2,{ q}_2^2,\phi,2M_{\rm x}\epsilon_{\rm x})\right]}/{{\cal N}_\phi \tau_{\rm hf}}}\,,
\end{array}
\end{equation}
\vskip -2mm
\noindent
where $ \phi$ is the angle between ${\bf q}_1$ and ${\bf q}_2$,
\begin{equation}\label{F}
{\cal F}(x,y,\phi,\beta)=\frac{xy\cos^2{\!\!\phi}\,
(\beta+x+y)}{\left[\beta^2+\beta(x+y)+
xy\cos^2{\!\phi}\right]^{3/2}}\,,\quad\mbox{and}
\end{equation}
\begin{equation}\label{tau_hf}
 \frac{1}{\tau_{\rm hf}}=\frac{5v_0M_{\rm x}\left(A_{\rm Ga}^2\!+\!A_{\rm As}^2\right)}{2 d\hbar\nu\,'l_B^2}
\end{equation}
[we have kept in ${\cal F}$ only terms nonvanishing after averaging over the ${\bf q}_1$ and ${\bf q}_2$ directions when  in Eq. \eqref{rate}].
The summation over $n$ in Eq. (\ref{rate}) has been performed for the case of unpolarized nuclei. In addition the  correlation length of the spatial nuclear momenta distribution has been considered to be smaller than the magnetic length $l_B$ and conventional width of the two-dimensional electron gas: $
  \,d\!\!=\!\!\left(\int|\chi(z)|^4dz\right)^{-1}\!
$.
(This value  is certainly not equal to
the quantum well width $d_{\rm QW}$, but constitutes a fraction of
it, e.g.: $d/d_{\rm QW}\simeq 1/3$.)

The rate $-d{\cal N}_{\rm x}/dt$ is
completely determined by Eqs. \eqref{distr1}-\eqref{tau_hf}. In the following calculations we use the following: (i) the kinematic and dynamic scattering fluxes simply add,
as independent contributions to the total rate; (ii) in the case of $T\!\ll\!\epsilon_{\rm Z}$ the  contribution to the rate due to the dynamic scattering relaxation flux is negligibly small; the same result is found if one of spin excitons in the initial state belongs to the thermodynamic condensate (i.e. ${\bf q}_1$ or/and ${\bf q}_2\in \{0\}$); (iii) $S({\bf q}_1,{\bf q}_2)$ does not depend on ${\bf q}_1$ and ${\bf q}_2$ for  kinematic scattering, and the summation in Eq. \eqref{rate}  reduces to
\begin{equation}\label{q_sum}
\sum_{{\bf q}_1,{\bf q}_2}[...]={\cal N}_{\rm x}^2-\sum_{{\bf q}_1,{\bf q}_2}N_{12}(1\!+\!N_{{\bf q}_1}\!+\!N_{{\bf q}_2})\,.
\end{equation}

In the $T\!\gapprox\!\epsilon_{\rm Z}$ region the spin-orbit relaxation channels are much more intense than the considered hyperfine coupling channel (see the next sections), and both spin-orbit and hyperfine coupling relaxation mechanisms compete with each other only in the $T\!\ll\!\epsilon_{\rm Z}$ case. Therefore we specifically study this situation. Then the dynamic spin-exciton - spin-exciton scattering is neglected, and the spin exciton creation term in Eq. (\ref{q_sum}) may be presented as $-\sum_{{\bf q}_1,{\bf q}_2}N_{12}(...)\approx -e^{-\mu/T}\sum_{{\bf q}_1,{\bf q}_2}N_{{\bf q}_1}N_{{\bf q}_2}(1\!+\!N_{{\bf q}_1}\!+\!N_{{\bf q}_2})/(1\!+\!N_{{\bf q}_1})(1\!+\!N_{{\bf q}_2})$. In the $\mu\gg T$ case this term is a negligible quantity  compared to ${\cal N}_{\rm x}^2$. If we consider $\mu\lapprox T$, then the term  is equal to $-{\cal N}_{\rm x}{\cal N}_{\rm x}^{(0)}$. So, if $T\ll\epsilon_{\rm Z}$, then for  any relation between  $\mu$ and $T$ one finds that Eq. \eqref{rate}  reduces to
\begin{equation}\label{rate2}
-d{n}_{\rm x}/dt={n}_{\rm x}\!\left[{n}_{\rm x}-{n}_{\rm x}^{(0)}\right]/2\tau_{\rm hf}\qquad(T\!\ll\!\epsilon_{\rm Z})
\end{equation}
[$n_{\rm x}(t)={\cal N}_{\rm x}(t)/{\cal N}_\phi$ and $n_{\rm x}^{(0)}={\cal N}_{\rm x}^{(0)}/{\cal N}_\phi$ to note the spin exciton concentrations].
In fact  under the conditions considered, the observable relaxation process is completed while still ${n}_{\rm x}^{(0)}t/2\tau_{\rm hf}\!\ll\! 1$, then
\begin{equation}\label{relaxation_law2}
n_{\rm x}(t)=\displaystyle{\frac{n_{\rm x}(0)}{1+n_{\rm x}(0)t/2\tau_{\rm hf}}}\,.
\end{equation}
This law is independent of the temperature but depends on the magnitude of the initial spin excitation $n_{\rm x}(0)$. The effective relaxation rate is $\sim n_{\rm x}(0)/2\tau_{\rm hf}\lapprox 0.1/\tau_{\rm hf}$ (if one assumes that $n_{\rm x}(0)\lapprox 0.1$).

\subsection{Spin exciton relaxation due to  hyperfine coupling together with the interaction of spin excitons with acoustic phonons}

In principle, the spin-exciton - phonon coupling mechanism  participates both in the spin-exciton - spin-exciton annihilation scattering and in the single-spin exciton one. However in the case of spin-exciton - spin-exciton scattering this relaxation channel represents only a small correction to those studied in the previous subsections, proportional to electron-lattice coupling constants.
Let us estimate the spin exciton-phonon relaxation governed by the single-exciton annihilation mechanism. We need to calculate the transition matrix element ${\cal M}_{\rm x-ph}$ between the state $|{\rm ket}\rangle\!=\!|{\rm SE},{\bf q}_1\rangle$ and some of final states $|{\rm bra}\rangle\!=\!{\hat P}_{{\bf k},s}^\dag|{\rm NST},n\rangle$ for the exciton-phonon operator determined by Eqs. \eqref{e-ph1} and \eqref{e-ph3}. Now the energy conservation law reads
${ E}_{\bf x}({\bf q}_1\!)=\hbar ck/l_B$,
where ${\bf k}\!=\!(k_z,{\bf q})$.
Meanwhile the ${{\bf q}}\!=\!0$ phonons do not contribute to the relaxation process, because action of the ${\cal H}_{\rm e-ph}(0)$ operator \eqref{e-ph3} on the $|{\rm SE},0\rangle$ state is reduced to multiplication by a constant -- hence ${\cal M}_{\rm x-ph}\!\equiv\!0$ due orthogonality of the $|{\rm SE},{\bf q}_1\rangle$ and $|{\rm NST},n\rangle$ states. If $q\!\not=\!0$ then the contribution to ${\cal M}_{\rm x-ph}$ is determined only by the first component of the  ket-state $|{\rm SE},{\bf q}_1\rangle$, namely by commutators  $\left[{\cal H}_{\rm e-ph}({\bf q}),{{\cal Q}^\dag_{{\bf
q}_1}}\right]|\{M\},0\rangle/{\sqrt{\nu\,'}}$. The latter according to Eq. \eqref{e-ph2} and commutation rules \eqref{commutators} vanish in case ${\bf q}_1\!=\!0$ being proportional to ${\bf q}\!\times\!{\bf q}_1$ at small $q_1$. This issue is a key point: the matrix element squared $|{\cal M}_{\rm x-ph}({\bf q}_1)|^2$ is proportional not only to the small constants of the hyperfine coupling and electron-phonon coupling but also to the temperature (more exactly to the small dimensionless parameter $M_{\rm x}T$). As a result, making  computations similar to those made above, we finally obtain  a relaxation rate linear in $n_{\rm x}$: $-d\Delta{n}_{\rm x}/dt\!=\!\Delta{n}_{\rm x}/\tau_{\rm hf\!-\!ph}$ [$\Delta{n}_{\rm x}$ to note the difference $n_{\rm x}\!-\!{n}_{\rm x}^{(0)}$], with the characteristic inverse time
\begin{equation}\label{tau_ph}
\frac{1}{\tau_{\rm hf\!-\!ph}}\sim\frac{\nu\,'v_0M_{\rm x}^3\left(A_{\rm Ga}^2\!+\!A_{\rm As}^2\right)\epsilon_{\rm Z}T}{\hbar cl_B^4dp_0^3\tau_D}\,.
\end{equation}
(under the considered conditions predominantly the deformation part of the $e$-phonon coupling contributes to the result). This value is much  smaller than the inverse time given by formula \eqref{tau_hf}. Much more important is comparison with another value governing also the single-spin exciton relaxation process related to phonon emission: namely, a certain characteristic
inverse time $1/\tau_{\rm so\!-\!ph}$ can be calculated in the case where spin non-conservation  instead of the hyperfine coupling is determined by the spin-orbit coupling.\cite{2di96,di99} It is found that at any parameters $1/\tau_{\rm so\!-\!ph}$ is much larger than $1/\tau_{\rm hf\!-\!ph}$ (by two or three orders of magnitude). We conclude that  spin exciton relaxation channels appearing due to the hyperfine coupling together with  electron-phonon coupling are very slow and may always  be neglected.

\section{The spin-orbit relaxation channels}

The spin-orbit relaxation channels, similarly to the hyperfine coupling mechanisms, may be subdivided into the two spin-exciton scattering channels and the single spin exciton ones. Among them there is a strong  spin-exciton - spin-exciton scattering process actually responsible for the spin exciton relaxation under the conditions of published experimental studies, \cite{zh93,fu08} namely at $T\sim 1\,$K and $B<10\,$T. This is the spin-exciton - spin-exciton dynamic scattering where the spin-flip is determined by the transition matrix element $\langle\mbox{fin}|{\hat H}_{\rm int}'|\mbox{ini}\rangle$ calculated for  operator \eqref{so_coul}, and states $|\mbox{ini}\rangle\!=\!{\cal Q}_{{\bf q}_1}^\dag{\cal Q}_{{\bf q}_2}^\dag|0\rangle/{\nu\,'}$ and $|\mbox{fin}\rangle\!=\!{\cal Q}_{{\bf q}}^\dag|0\rangle/\sqrt{{\nu\,'}}$. Being constrained   by  energy  $E({\bf q}_1)\!+\!E({\bf q}_2)\!=\!E({\bf q})$ and momentum conservation ${\bf q}_1\!+\!{\bf q}_2\!=\!{\bf q}$, this process  occurs if ${\bf q}_1{\bf q}_2\!=\!\epsilon_{\rm Z}M_{\rm x}$; i.e. the phase volume of the scattered spin excitons is essentially restricted. In particular, if the scattering spin excitons belong to the thermodynamic condensate, this relaxation mechanism is switched off. In fact the dynamic relaxation channel works well only when $T\gapprox \epsilon_{\rm Z}$, giving the  relaxation time $\sim\!\!10\,$ns.\cite{di99,di09,foot2} However if $T\ll\epsilon_{\rm Z}$, the characteristic time is drastically extended, as it is  proportional to the double exponent $\sim\!e^{2\epsilon_{\rm Z}/T}$ (see Ref. \onlinecite{di99}).
Therefore, studying exactly the $T\ll\epsilon_{\rm Z}$ case where the spin-orbit and hyperfine coupling relaxations are competing, we consider the spin-exciton - spin-exciton kinematic processes provide more intense relaxation. In  the excitonic representation these are determined by operators  \eqref{so_e-ph3} and \eqref{srp} which do not conserve the number of spin excitons.

\subsection{Relaxation via a smooth random potential}

The spin-orbit relaxation channel in presence of a  smooth random potential, is again governed by the kinetic equation \eqref{rate} where $S({\bf q}_1,{\bf q}_2)=(2\pi/\hbar)\sum_{\bf q}\left|{\cal M}_{if}^{\rm srp}({\bf q}_1,{\bf q}_2,{\bf q})\right|^2\delta({\cal E}_{q}\!-{\cal E}_{{ q}_1}\!-{\cal E}_{{ q}_2}\!-\epsilon_{\rm Z})$, where
${\cal M}_{if}^{\rm srp}=\langle\mbox{fin}|{\hat \varphi}'|\mbox{ini}\rangle$ with initial and final states $|\mbox{ini}\rangle={\cal Q}_{{\bf q}_1}^\dag{\cal Q}_{{\bf q}_2}^\dag|0\rangle/{\nu\,'}$ and $|\mbox{fin}\rangle\!=\!{\cal Q}_{{\bf q}}^\dag|0\rangle/\sqrt{{\nu\,'}}$ respectively.
Taking into account that ${\cal E}_{{ q}_1},{\cal E}_{{ q}_2}\ll\epsilon_{\rm Z}$, the argument of the $\delta$-function may be set ${\cal E}_{q}-\epsilon_{\rm Z}$, and using Eq. (C.1), we obtain the squared matrix element  $\left|{\cal
M}_{if}^{\rm srp}\right|^2\!=\!
2(u^2\!+\!v^2)
  \left|q^*\overline{\varphi}({\bf q}^*)\right|^2/{\nu\,'\!N_{\phi}^2}$,
where
$q^*=\sqrt{2M_{\rm x}\epsilon_{\rm Z}}$, and the scattering probability
independent of ${\bf q}_1$  and ${\bf q}_2$: $S=1/\!{\cal N}_\phi\tau_{\rm so}^{\rm srp}$. The characteristic inverse relaxation time
is
\begin{equation}\label{tau_srp}
1/\tau_{\rm so}^{\rm srp}= 16\pi^2(u^2+v^2)M_{\rm x}^2\epsilon_{\rm Z}{\overline K}(q^*)/\nu\,'\hbar\,.
\end{equation}
Here ${\overline K}(q)$ stands for the Fourier component of the
correlator. If the latter represents a Gaussian function (see Sec. II-D), then ${\overline K}(q^*)=\pi\Delta^2\exp{(-M_{\rm x}\epsilon_{\rm Z}\Lambda^2/2l_B^2)}$.\cite{foot3} We note  that it  depends exponentially on the magnetic field squared: $\sim\!e^{-\gamma B^2}$ (the spin exciton mass is assumed to be independent of $B$).
As mentioned earlier, this time is assumed much longer than the times of thermalization and therefore determines the relaxation while the irreversibility occurs due to the fast thermalization. The relaxation rate can then be calculated as in Sec. III-C. Whether or not  the thermodynamic condensate exists, it is governed by equation
\begin{equation}\label{rate_srp}
-d{n}_{\rm x}/dt={n}_{\rm x}\!\left[{n}_{\rm x}-{n}_{\rm x}^{(0)}\right]/2\tau_{\rm so}^{\rm srp}
\end{equation}
differing from Eq. \eqref{rate2} only by the replacement of  $\tau_{\rm hf}$ with $\tau_{\rm so}^{\rm srp}$. Likewise one obtains Eq. \eqref{relaxation_law2} with the same substitution.

\subsection{Electron-phonon coupling mechanism of the dissipation}

We study in this subsection the spin-exciton - spin-exciton scattering process, where there are two spin excitons in the initial state and a single spin exciton plus an emitted phonon in the final state. (For a discussion of single spin-exciton annihilation due to phonon emission, see comments at the end of Sec. III-D). In this case the conservation laws read:
\begin{equation}\label{Energi+momentum}
\begin{array}{l}
{\bf q}_1+{\bf q}_2={\bf q}+{\bf q}_{\rm ph}  \quad\mbox{and}\\
E({\bf q}_1)+E({\bf q}_2)=E({\bf q})+\hbar c\sqrt{k_z^2+{q}_{\rm ph}^2}.
\end{array}
\end{equation}
Now the kinetic equation for annihilated and created spin excitons is
\begin{equation}\label{rate_SE-SE+PH}
-\frac{d{\cal N}_{\bf x}}{dt}\!=\!\frac{1}{2}\sum_{{\bf q}_{1},{\bf q}_2,{\bf q}}S({\bf q}_1,{\bf q}_2,{\bf q})
  \left[N_{{\bf q}_1}N_{{\bf q}_2}\left(1+N_{{\bf q}}+N_{\rm ph}\right)-N_{{\bf q}}N_{\rm ph}\left(1+N_{{\bf q}_1}+N_{{\bf q}_2}\right)
  \right],
\end{equation}
Due to the $T\ll\epsilon_{\rm Z}$ condition we can neglect values ${\cal E}_{q_1}$ and ${\cal E}_{q_2 }$ in the $E({\bf q}_1)\!+\!E({\bf q}_2)\!-\!E({\bf q})\!-\!\varepsilon_{\rm ph}$ argument of the $\delta$-function when calculating the scattering probability, therefore
\begin{equation}\label{prob}
S({\bf q}_1,{\bf q}_2,{\bf q})=\frac{2\pi}{\hbar}\sum_{k_z,{\bf q}_{\rm ph},s}|{\cal M}_{\rm x-ph}({\bf q}_1,{\bf q}_2,{\bf q},k_z,{\bf q}_{\rm ph},s)|^2\delta(\epsilon_{\rm Z}\!-\!{\cal E}_{q}\!-\!\hbar c\sqrt{k_z^2+{q}_{\rm ph}^2}).
\end{equation}
The matrix element is ${\cal M}_{\rm x-ph}\!=\!\langle{\rm fin}|{\hat H}_{\rm e-ph}|{\rm ini}\rangle$, where the electron-phonon Hamiltonian is presented by Eqs. \eqref{e-ph1}-\eqref{vertex} (with change from ${\cal H}_{\rm e-ph}$ to ${\cal H}'_{\rm e-ph}$, see Eq. \eqref{so_e-ph3}), and bra- and ket-vectors are $|{\rm fin}\rangle\!=\!{\hat P}^\dag_{k_z,{\bf q}_{\rm ph},s}{\cal Q}_{{\bf q}}^\dag|0\rangle/\sqrt{{\nu\,'}}$ and $|\mbox{ini}\rangle={\cal Q}_{{\bf q}_1}^\dag{\cal Q}_{{\bf q}_2}^\dag|0\rangle/{\nu\,'}$ respectively. Using Eq. (C.1) we keep again only terms $\sim \nu\,'\!/{\cal N}_\phi$ contributing to the result in the leading approximation. Finally, by doing in the spirit of manipulations above, one obtains the relaxation rate \eqref{rate_SE-SE+PH} in the form
\begin{equation}\label{so_e-ph_rate}
-d{n}_{\rm x}/dt={n}_{\rm x}\![{n}_{\rm x}\!-{n}_{\rm x}^{(0)}]/2\tau_{\rm so}^{\rm e-ph}
\end{equation}
similar to Eqs. \eqref{rate2} and \eqref{rate_srp}. Now the temperature-independent constant characterizing the rate is$\;$\cite{foot4}
\begin{equation}\label{tau_e-ph}
\displaystyle{1/\tau_{\rm so}^{\rm e-ph}=\frac{4(u^2\!+\!v^2)M_{\rm x}^2\epsilon_{\rm Z}^3{\cal G}({M}_{\rm x}c^2\hbar^2/\epsilon_{\rm Z}l_B^2)}{\nu\,'\!c\hbar l_B^2p_0^3\tau_D}}\,,
\end{equation}
where $$
{\cal G}(\xi)\!=\!\int_0^{x_0(\xi)}\!\!\!\!\!\!dx(x\!-\!x^2)/\sqrt{1\!-\!2\xi x/(1\!-\!x)^2}
$$
[$x_0\!=\!1\!+\!\xi\!-\!\sqrt{\xi^2\!+\!2\xi}$]. In  the derivation we have set $1/\tau_A\approx 1/\tau_D$, because estimation shows that contribution of the deformation coupling is dominating the polarization one under the considered conditions (cf. Sec. III-D). Unlike the characteristic value \eqref{tau_srp} which decreases exponentially with the magnetic field, the inverse time \eqref{tau_e-ph} grows  and depends on $B$ with  the power law $\sim B^3$. This increase comes from the $\epsilon_{\rm Z}^3$ factor in equation \eqref{tau_e-ph}, which reflects the increased phase space available from the emission of phonons at high fields.\cite{2di96}

\section{Comparison of the hyperfine coupling and spin-orbit relaxation channels. Discussion.}

Summing up the right-hand sides of Eqs. \eqref{rate2}, \eqref{rate_srp}, and \eqref{so_e-ph_rate}, we find the total relaxation flux:
\begin{equation}\label{total}
-dn_{\rm x}/dt=\left(n_{\rm x}-n_{\rm x}^{(0)}\right)\left[\frac{n_{\rm x}}{2}\left(\frac{1}{\tau_{\rm hf}}+\frac{1}{\tau_{\rm so}^{\rm srp}}+\frac{1}{\tau_{\rm so}^{\rm e-ph}}\right)\right]\,.
\end{equation}
As the inverse relaxation time is in fact proportional to $n_{\rm x}$, we characterize  the relaxation process at a
substantial initial excitation $n_{\rm x}(0)$. The latter value experimentally is $\sim 0.1$ and under the assumed conditions $T\lapprox 0.1\,$K and $B> 10\,$T (where the equilibrium concentration $n_{\rm x}^{(0)}\!<\! 10^{-4}$) one finds the law $n_{\rm x}(t)\!=\!n_{\rm x}(0)/[1\!+\!n_{\rm x}(0)t/2\tau_{\rm tot}]$, where
\begin{equation}\label{tot}
\frac{1}{\tau_{\rm tot}}=\frac{1}{\tau_{\rm hf}}+\frac{1}{\tau_{\rm so}^{\rm srp}}+\frac{1}{\tau_{\rm so}^{\rm e-ph}}.
\end{equation}

Estimates of the $\tau_{...}^{...}$-values are possible if we specify material parameters included in formulae \eqref{tau_hf},\eqref{tau_srp} and \eqref{tau_e-ph}. Some of them have been already given in Secs. II-B and II-C and in Appendixes A and B. In addition we consider $c=5\cdot10^5\,$cm/s and $\epsilon_Z\!=\!\!0.0255B\,$meV. Other parameters related to modern wide quantum-well structures could be chosen as $u^2\!+v^2\!=\!10^{-3}/B$, ,
$\Lambda\!=\!50\,$nm, $\Delta\!=\!0.3\,$meV, and $d\!=\!8.1\,$nm
(here $B$ is assumed to be measured in Teslas; cf. also estimates
in Ref. \onlinecite{di09}). However, estimate of the effective
spin-exciton mass $M_{\rm x}$ strongly depends on the
finite thickness form-factor. There are experimental data
where $M_{\rm x}$ is found at comparatively low magnetic fields:
(i) $1/M_{\rm x}\!\approx\!1.2\,$meV at $B\!=\!2.27\,$T and
$\nu\!=\!1$ in the $33\,$nm quantum well;\cite{ga08} (ii)
$1/M_{\rm x}\!\approx\!1.51\,$meV at $B\!=\!2.69\,$T and
$\nu\!=\!1$ in the $23\,$nm quantum well;\cite{kukush09} and
(iii) $1/M_{\rm x}\!\approx\!0.44\,$meV at $B\!=\!2.9\,$T and
$\nu\!=\!1/3$ in the $25\,$nm quantum well.\cite{kukush06} For
these fields characterized by the inequality $l_B\!>\!d$, the
$B$-dependence should be approximately $1/M_{\rm x}\sim B^{1/2}$,
but in the $l_B\!<\!d$ strong field regime the inverse mass grows
much more weakly with $B$. Based on these data, the semi-empirical
analysis using characteristic GaAs/AlGaAs form factors allows us to
consider values $1/M_{\rm x}\!\simeq 2\,$meV at $\nu\!=\!1$ and
$1/M_{\rm x}\!\simeq 0.7\,$meV at $\nu\!=\!1/3$ as the
characteristic ones for the $10\,$T$<\!B\!<\!25\,$T range. (Note
that at a given field $B$ the estimate $M_{\rm
x}^{-1}|_{\nu\ \!\!<\!\!\ 1}\!\simeq\!\nu{\,}'\!\cdot\!M_{\rm
x}^{-1}|_{\nu\!=\!1}$ holds according to the semi-phenomenological
theory.\cite{lo93})

Numerical values of the characteristic inverse relaxation times are plotted in Fig.1. as a function of  magnetic field. We remark that actual times should be  longer by factor $\sim 2/n_{\rm x}(0)\!\sim 20-50$ because of the  non-exponential solution of equation \eqref{tot}. The $B$-dependence of the relaxation rate is  non-monotonic. In the region $10\,$T${}\!<\!B\!<\!30\,$T the relaxation regime  switches twice  between the spin-orbit and hyperfine coupling dominance, taking  maxima $\simeq 18\,$T and $\simeq 12\,$T in the $\nu\!=\!1$ and $\nu\!=\!1/3$ cases respectively.
The reason that the hyperfine interaction becomes dominant is that for increasing magnetic field the nuclei remain disordered, while the   random potential is  effectively smoothed by the cyclotron motion. At very high fields the spin-orbit interaction again dominates because of the
increasing phase-space for the emission of phonons. On the basis of these estimates we conclude that the hyperfine coupling relaxation channel should be dominant approximately from $16$ to $29\,$T in the $\nu=1$ quantum Hall ferromagnet and from $11$ to $24\,$T for  $\nu=1/3$. The latter case would seem to be more  accessible  to the  experimental study of  the hyperfine coupling relaxation mechanism, because usual electron concentrations in GaAs structures do not allow one to attain  fields stronger $10\,$T in the  $\nu\!=\!1$ quantum Hall system. We note a feature of the hyperfine coupling relaxation: its rate is vanishing in the case of spin-polarized nuclei. This  should distinguish  the hyperfine coupling mechanism from that of spin-orbit and provide a test of the theory. If the nuclear spins could be fully polarized, then only spin-orbit relaxation would be important and there should be crossover between the regime limited by the random potential
and the very high field regime of phonon emission.We emphasize also that our results should be valid in immediate vicinity of 1 or 1/3 fillings. Recent experiments show that if  $\nu$ differs by more than about  0.1 from these special values,  one observes
a two-mode spectrum of spin excitations -- above and below the Zeeman gap.\cite{dr10} Interaction of these two types of spin waves could considerably accelerate the relaxation.

In conclusion, we have reported on a new spin relaxation mechanism in a spin polarized strongly correlated two-dimensional electron gas that appears at low temperatures and in strong magnetic fields. This mechanism is related only to the hyperfine coupling with GaAs nuclei and no other interactions are needed for this relaxation channel. The full calculation of relaxation  displays a competition of the hyperfine coupling and
spin-orbit relaxation processes, which can be summarized by equations \eqref{tau_hf},\eqref{tau_srp}, and \eqref{tau_e-ph}. Under the assumed conditions the relaxation process occurs non-exponentially with time.  The rate does not depend on temperature but  depends on the magnetic field non-monotonically as can be seen in Figure 1, which is plotted using estimated material and device parameters taken from experiment.
The estimate of the hyperfine relaxation  depends on the assumed randomness of the nuclear spins and a test of the theory would be to polarize the nuclear spins.

S.D. thanks the Russian Fund of Basic Research and the LIA Condensed Matter and Theoretical Physics Program (ex ENS-Landau) for support, and the Laue Langevin
Institute (Grenoble) for hospitality.

\appendix

\section{Calculation of the hyperfine coupling parameters $A_{\rm Ga}$ and $A_{\rm
As}$}

 We proceed from formula $A_n\!=\!(16\pi\mu_B\mu_n/3I_n)|u(\mbox{\boldmath $R$}_n)|^2\!,\,\,$\cite{dy84,me02} where $\mu_n$ is the nuclear magnetic moment, and $u(\mbox{\boldmath $R$}_n)$ is the conduction electron Bloch function at the nucleus. $u(\mbox{\boldmath $R$})$ is assumed to be normalized as $\displaystyle{{\int}}\!|u(\mbox{\boldmath $R$})|^2d^3\!R\!=\!1$,
 where the integration is performed within the GaAs two atom unit cell having volume $v_0\!=\!45.2\,$\AA${}^3$. It seems to be the only estimations of $|u(\mbox{\boldmath $R$}_{\rm Ga})|^2$ and $|u(\mbox{\boldmath $R$}_{\rm As})|^2$ were done in Ref. \onlinecite{pa77} and subsequently cited by other authors (cf. Ref. \onlinecite{dy84}). Using these and the $\mu_n$ values for As and for the Ga${}^{69}$ and Ga${}^{71}$ stable isotopes: $\mu_{\rm As}\!=\!1.44$, $\mu_{{\rm Ga}^{69}}\!=\!2.017$ and $\mu_{{\rm Ga}^{71}}\!=\!2.56$ (in units of the nuclear magneton $\mu_N\!=\!3.15\cdot10^{-9}\,{\rm meV/G}$),\cite{ra89} we find $A_{{\rm Ga}^{69}}\simeq0.038\,$meV, $A_{{\rm Ga}^{71}}\simeq0.049\,$meV and $A_{\rm As}\simeq0.046\,$meV. Ratio of the Ga${}^{69}$ and Ga${}^{71}$ amounts in the semiconductor is considered to be equal to 3:2, therefore the result is $$\sum_{{\rm within}\:{\rm unit}\:{\rm cell}}\!\!\!\!\!A_n^2=0.6\left(A_{{\rm Ga}^{69}}\right)^2\!+0.4\left(A_{{\rm Ga}^{71}}\right)^2\!+\left(A_{{\rm As}^{71}}\right)^2\!\approx 4\cdot10^{-3}\,\mbox{meV}^2\,.  \eqno (\mbox{A}.1)
$$


\section{acoustic phonon life times  $\mbox{\large
\boldmath $\tau$}_{\!\mbox{$l$}}$ and $\mbox{\large
\boldmath $\tau$}_{\!\mbox{$t$}}$}

If we take ${\hat x},\;{\hat y},\;{\hat z}$ to be the
directions of the principal crystal axes, then for longitudinal phonons we
obtain${\,}$\cite{2di96,di00}
$$
  \frac{1}{\tau_{l}({\bf k})}=
  \frac{1}{\tau_D} +
  \frac{45p_0^2}{k^8\tau_P}q_x^2q_y^2k_z^2\,,     \eqno (\mbox{B}.1)
$$
where
$$
  \tau_D^{-1}=\frac{{\Xi}^2p_0^3}{2\pi \hbar \rho c^2},\quad \tau_P^{-1}=
  \left(\frac{ee_{14}}{\kappa}\right)^2\frac{8\pi p_0}
  {5\hbar \rho c^2}\eqno (\mbox{B}.2)
$$
(${\bf q}$ and $k_z$ in this Appendix are considered to have common dimension.) Transverse phonons in a cubic crystal do not induce a deformation field.\cite{gale87}
Actually, we need only the inverse time $1/{\tau_t}$ averaged over all directions of the
transverse
phonon polarization. If the transverse phonon distribution satisfies the condition that their
polarizations are equiprobable, then for either of the two polarization the averaging
yields$\,$\cite{2di96,di00}
$$
  \overline{{\tau_{{t}}}^{-1}}=\frac{5p_0^2}
  {2k^6\tau_P}\left(
  q_x^2q_y^2+q^2k_z^2-\frac{9q_x^2q_y^2k_z^2}{k^2}\right)\,.
                                                         \eqno (\mbox{B}.3)
$$
We have used in Eqs. (B.2) and (B.3) common notations: $\Xi\simeq 17.5\,$eV and $e_{14}\simeq
-0.16\,$C/m${}^2$ are the relevant deformation potential and piezoelectric constant of the
GaAs crystal,
$\rho\approx 5.3$\,g/cm${}^2$ is the GaAs density, $\kappa\approx 12.85$ is the dielectric
constant. As a result, we find $\tau_D\simeq 0.8\,$ps and $\tau_P\simeq 35\,$ps.
\vspace{-2mm}


\section{}
The four-${\cal Q}$-operator expectation value is calculated with the help of Eq. (\ref{commutators})
$$
\begin{array}{l}\left\langle 0|{\cal Q}_{{{\bf q}_2}'}{\cal Q}_{{{\bf q}_1}'}{\cal Q}_{{\bf
q}_{\!1}}^{\dag}\!{\cal Q}_{{\bf q}_2}^{\dag}|\,0\right\rangle\\
\qquad{}\qquad\displaystyle{=\delta_{{\bf q}_1\!+\!{\bf
q}_2\!,\,{{\bf q}_1}'\!+\!{{\bf q}_2}'}\left[e^{i\phi}\left(\left\langle 0\left|{\cal A}_{{{\bf q}_2}'\!-\!{{\bf q}_2}}{\cal A}^\dag_{{{\bf q}_1}\!-\!{{\bf q}_1}'}\right|0\right\rangle-\frac{\nu\,'}{{\cal N}_\phi}\right)\right.}\\
\quad{}\qquad{}\qquad{}\qquad+\displaystyle{\left.e^{-i\phi}\left(\left\langle 0\left|{\cal A}_{{{\bf q}_2}'\!-\!{{\bf q}_1}}{\cal A}^\dag_{{{\bf q}_2}\!-\!{{\bf q}_1}'}\right|0\right\rangle-\frac{\nu\,'}{{\cal N}_\phi}\right)\right]}\,,
\end{array}   \eqno (\mbox{C}.1)
$$
where $\phi=\left({{\bf q}_1}'\!\times\!{{\bf q}_1}\!+\!{{\bf q}_2}'\!\times\!{{\bf q}_2}\right)_z/{2}$, and $\nu\,'$ is considered to be equal $\nu$ if $\nu\leq 1$, or 1 if the filling factor is integer.
In the important case of integer $\nu$: $\langle 0|{\cal A}_{\bf q}{\cal A}_{\bf q}^\dag|0\rangle=\delta_{{\bf q},0}$. Then, if calculating the matrix element (\ref{M_kin}), the $\langle 0|{\cal A}_{...}{\cal A}_{...}^\dag|0\rangle$ terms do not contribute to the probability transition (\ref{FGR}) owing to the energy conservation condition (\ref{conservation}). This means that the kinematic scattering would be determined only by the double-commutation expectation value $\left\langle {\rm fin}\left|\left[\left[{\hat H}_{\rm hf},{\cal Q}_{{\bf
q}_{\!1}}^{\dag}\right],{\cal Q}_{{\bf q}_2}^{\dag}\right]\right|0\right\rangle$ similar to the case of the dynamic scattering, cf. Eq. \eqref{M_dyn1}.

In Eq. (\ref{eigen2}) the action of
the double-commutation term  leads to the state
$$
\left[[{\hat H}_{\rm int},{\cal Q}_{{\bf q}_1}^{\dag}],
{\cal Q}_{{\bf q}_2}^{\dag}\right]|0\rangle=\frac{4}{{\cal N}_\phi}\sum_{{\bf q}'}W(q')\sin{\left(\frac{{\bf q}'\!\times\!{\bf q}_1}{2}\right)}\!\cdot\!\sin{\left(\frac{{\bf q}'\!\times\!{\bf q}_2}{2}\right)}{\cal Q}_{{\bf
q}_{\!1}\!-\!{\bf q}'}^{\dag}\!{\cal Q}_{{\bf q}_2\!+\!{\bf q}'}^{\dag}|0\rangle\,.    \eqno(\mbox{C}.2)
$$

\begin{figure}[h]
\begin{center} \vspace{5.mm}
\hspace{-10.mm}
\includegraphics*[width=1.\textwidth]{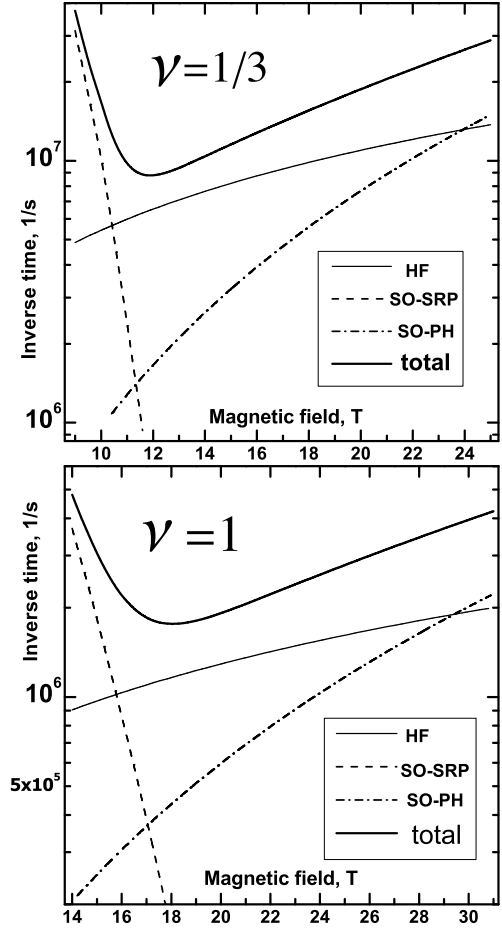}
\end{center}
\vspace{-10.mm}
 \caption{Calculated inverse relaxation times as a function of magnetic field  $B$ from  formulae \eqref{tau_hf}, \eqref{tau_srp} and \eqref{tau_e-ph} corresponding to hyperfine, $1/\tau_{\rm hf}$ (solid line), spin-orbit with random potential $1/\tau_{\rm so}^{\rm srp}$ (dash), and spin-orbit with phonon emission $1/\tau_{\rm so}^{ph}$ (dash-dot), respectively.
 Specific material parameters are given in the text.  The bold solid line is the
 calculated combined inverse time \eqref{tot}.}
\end{figure}


\begin{thebibliography}{99}
\bibitem{foot2}
{This calculation is in satisfactory  agreement with experimental data of refs. \onlinecite{fu08}, where the spin relaxation has
been studied if the initial deviation  was definitely an excited spin-exciton state.
 We note that those authors refer to the publication \onlinecite{di04}, which treated Goldstone mode relaxation, but references \onlinecite{di99} and \onlinecite{di09} would have been more appropriate.}
\bibitem{foot3}
{Formula \eqref{tau_srp} was derived earlier by using a more general approach, wherein the Goldstone mode relaxation was considered.\cite{di04} However, during the Goldstone mode relaxation the thermodynamic condensate  sequentially increases and decays, and its evolution is also governed by
the time \eqref{tau_srp}.}
\end{thebibliography}
\end{document}